\documentclass[twocolumn,showpacs,groupedaddress,preprintnumbers]{revtex4-1}

\usepackage{graphicx}  
\usepackage[caption=false]{subfig}
\usepackage{dcolumn}   
\usepackage{bm}        
\usepackage{amssymb}   
\usepackage{braket}
\usepackage{amsmath}
\usepackage{color}  	

\graphicspath{{jaxodraw/}}

\usepackage{def}

\newcommand{\beq}{\begin{equation}}
\newcommand{\eeq}{\end{equation}}

\begin{document}

\title{Amplitude Factorization in the Electroweak Standard Model}
\author{Simon Pl\"atzer}
\affiliation{Institute of Physics, NAWI Graz, University of Graz, Universit\"atsplatz 5, A-8010 Graz, Austria}
\affiliation{Particle Physics, Faculty of Physics, University of Vienna, Boltzmanngasse 5, A-1090 Wien, Austria}
\affiliation{Erwin Schr\"odinger Institute for Mathematics and Physics, University of Vienna, A-1090 Wien, Austria}
\author{Malin Sj\"odahl}
\affiliation{Department of Physics, Lund University, Box 118, 221 00 Lund, Sweden}
\affiliation{Erwin Schr\"odinger Institute for Mathematics and Physics, University of Vienna, A-1090 Wien, Austria}
\date{\today}

\begin{abstract}
  We lay out the basis of factorization at the amplitude level for
  processes involving the entire Standard Model. The factorization
  appears in a generalized eikonal approximation in which we expand
  around a quasi-soft limit for massive gauge bosons, fermions, and
  scalars. We use the chirality-flow formalism to express loop
  exchanges or emissions as operators on chiral structures. This forms
  the basis for amplitude evolution with parton exchange and branching
  in the full Standard Model, including the electroweak sector.
\end{abstract}

\maketitle

\section{Introduction}

The measurements and searches for new physics at current and future
colliders operate through observables which resolve widely different
energy scales between the hard scattering and the details of the
observed final state. Such observables, which we generally refer to as
infrared sensitive, are computable in perturbation theory, however,
the appearance of large logarithms of the scale ratios and other
resolution parameters invalidates the truncation of the perturbative
series at any fixed order in the (small) coupling parameter. Instead,
resummation --- which is unavoidably tied to the description of
multiple emissions and properties of large multiplicity final states
--- is required to capture the physical behavior.

Resummation is only possible if factorization takes place, {\it i.e.}
when we can build up scattering amplitudes with many emissions and
exchanges of the interacting particles from repeated simple and
universal building blocks. This is well understood in the context of
the strong interaction (see {\it e.g.} \cite{Catani:1999ss}), and has
paved the way for the description of jets, and ultimately the
development of versatile simulations of high energy collisions (see
{\it e.g.} \cite{Forshaw:2020wrq}). While the strong interaction,
described by Quantum Chromodynamics (QCD), contributes the bulk of the
complexity in hadronic final states, at high enough energies there is
no kinematic suppression mechanism for electroweak interactions. All
Standard Model degrees of freedom need to be taken into account to
reliably predict the details of the final states in which we strive to
observe deviations from the Standard Model at colliders.

In this paper, we outline a formalism which serves as the basis for
generalizing soft gluon evolution
\cite{Kidonakis:1998nf,Contopanagos_1997,Oderda:1999kr,Sjodahl:2009wx,Platzer:2013fha,AngelesMartinez:2018cfz,DeAngelis:2020rvq,Platzer:2020lbr}
and amplitude level parton branching
\cite{Forshaw:2019ver,Loschner:2021keu}. 
Our formalism also accounts for
the exchange and emission of electroweak bosons, along with additional
effects from the electroweak interaction of the fermions in the
Standard Model. This will allow us to describe observables which are
sensitive to changes in the isospin composition of emitting systems,
as well as to account for the chiral nature of the electroweak
interactions. Our work will provide the fundamental building blocks to
apply and extend amplitude level evolution to the resummation of
electroweak effects
\cite{Ciafaloni:1998xg,Denner:2000jv,Denner:2001gw,Ciafaloni:2005fm,Chiu:2007yn,Manohar:2018kfx}.
It will also provide a thorough framework for the construction of
parton branching algorithms which coherently treat electroweak and QCD
effects on equal footing. Our framework will complement existing
approaches of electroweak showers
\cite{Kleiss:2020rcg,Masouminia:2021kne,Brooks:2021kji} which are
based on emission amplitudes only \cite{Chen:2016wkt} and will link to
electroweak evolution for strictly high energies in the
quasi-collinear limit, {\it e.g.}
\cite{Bauer:2017isx,Bauer:2018xag}. It is also in shape to include the
effects of mixing, decays and the projection onto observed states as
outlined in \cite{Platzer:2022jny}. The latter can lead to
possible miscancellations of logarithmically enhanced contributions
\cite{Manohar:2014vxa} or to mechanisms restoring the cancellation
\cite{Reiner:2021bol}.

We focus in particular on the underlying kinematic region in which
factorization, including different mass shells and possibly recoils,
take place, and on the structure of the evolving amplitude as a vector
in the space of isospin and chiral structures, similar to how it is
typically described as a vector in color space.  This will form the
basis of an actual implementation of an evolution algorithm within
amplitude evolution frameworks such as the \texttt{CVolver}
\cite{Platzer:2013fha,DeAngelis:2020rvq} library.

We first set the notation in \secref{sec:notation}.
The basis of the factorization of fully massive amplitudes is
described in \secref{sec:factorization}, with the kinematics
facilitating the factorization detailed in \secref{sec:kinematics}.
Self-energy insertions, wave function renormalization and cutting of
unresolved lines is discussed in \secref{sec:self-energy}, and in
\secref{sec:flow} the factorization is complemented with a complete
flow picture, entailing chiral and isospin structures.  Finally we
conclude in \secref{sec:conclusion}.

\section{Notation}
\label{sec:notation}

Central to our analysis is the formalism of treating amplitudes as
abstract vectors in a space of tensor structures of (internal) quantum
numbers, {\it i.e.} color, isospin and spinor indices. In order to
set the notation for this we start with a simple example: Consider
$q(1)Q(2)\to q(3)Q(4)$ scattering via a gluon exchange, then the
amplitude would be written as 
\begin{multline}
  {\cal M} = G\ u_{i_1,I_1,\alpha}(p_1) u_{i_2,I_2,\gamma}(p_2)
  \bar{u}^{i_3,I_3}_\beta(p_3) \bar{u}^{i_4,I_4}_\delta(p_4) \times \\
  \left(t^a\right)^{i_1} {}_{i_3}\left(t^a\right)^{i_2} {}_{i_4}\
  \delta^{I_1}_{I_3} \delta^{I_2}_{I_4}\
  \left(\overline{P_{c_1}}\gamma^\mu P_{c_3}\right)_{\beta \alpha}
  \left(\overline{P_{c_2}}\gamma_\mu P_{c_4}\right)_{\delta \gamma} \ ,
  \label{eq:amp}
\end{multline}
where $i_{1..4}$ are the color indices, $I_{1..4}$ are the isospin
indices, and $c_{1..4}$ are the chiralities we obtain after applying
chiral projectors $P_c$ to the external quark lines, and
$\alpha,\beta,\gamma,\delta$ are spinor indices. $G$ collectively
denotes all other coupling, symmetry factors and propagator
denominators. The external wave functions carry explicit momenta, and
will specify explicit quantum numbers like spin to be measured (which
we have suppressed for readability). We will write this amplitude as
\begin{equation}
  | {\cal M}\rangle = G\ |T\rangle \ ,\qquad   {\cal M} = \langle \psi | {\cal M}\rangle
\end{equation}
such that the tensor structure is contained in $|T\rangle$, and
explicit color, isospin and chirality is carried by
\begin{multline}
  \langle \psi | =
  u_{i_1,I_1,\alpha}(p_1) u_{i_2,I_2,\gamma}(p_2)
  \bar{u}^{i_3,I_3}_\beta(p_3) \bar{u}^{i_4,I_4}_\delta(p_4)\\
  \langle \{i_1,...\},\{I_1,...\},\{\alpha,...\} | \ ,
\end{multline}
which signifies an abstract version of the external wave function. The
components of the tensor $T$ in terms of explicit indices are
recovered by
\begin{multline}
  \langle \{i_1,...\},\{I_1,...\},\{\alpha,...\} |T\rangle =\\
    \left(t^a\right)^{i_1} {}_{i_3}\left(t^a\right)^{i_2} {}_{i_4}\
  \delta^{I_1}_{I_3} \delta^{I_2}_{I_4}\
  \left(\overline{P_{c_1}}\gamma^\mu P_{c_3}\right)_{\beta \alpha}
  \left(\overline{P_{c_2}}\gamma_\mu P_{c_4}\right)_{\delta \gamma} \ .
\end{multline}

\section{Strategy of factorization}
\label{sec:factorization}

We consider a subset of diagrams (which we label by the symbolic index
$s$) for a certain process in which $m$ lines with unresolved
particles of flavors $\{g_i\}_m$ carry `soft' momenta $\{k_i\}_m$, and
are emitted from, or exchanged in-between, a subset $h_s$ of the $n$
other external lines of flavors $\{f_i\}_n$, which carry `hard'
momenta $\{q_i\}_n$. The subdiagrams involving the emissions and
exchanges will then attach to an amplitude with on- or off-shell lines
of flavor $\{f'_i\}_n$, which carry momenta $P_{i,s} = q_i + K_{i,s}$,
with $K_{i,s}$ being some linear combination of the emitted and
exchanged momenta if $i\in h_s$, and $P_{i,s}=q_i$ if $i\notin h_s$ is
an external line not connecting to an unresolved line.

Having singled out a certain subgraph from the amplitude as described
above, we can write it as
\begin{eqnarray}
  \label{eq:offshell} 
  | && {\cal M}_{\{f_i\}_n,\{g_i\}_m}(\{q_i\}_n;\{k_i\}_m)\rangle
  \nonumber \\
  &&= \sum_{\{f'_i\}_n}\sum_s
            {\mathbf
              R}_{s;\{f_i\}_n,\{g_i\}_m}^{\{f'_i\}_n}(\{q_i\}_n;\{k_i\}_m) \\
            && \prod_{i\in h_s}\frac{{\mathbf P}_i(q_i+K_{i,s},M_i) }{(q_{i}+K_{i,s})^2-M_{i}^2}
            |{\cal M}_{\{f'_i\}_n} (\{P_i\}_n)\rangle + ...  \nonumber
\end{eqnarray}
in which ${\mathbf P}_i$ represents the propagator numerator of the
hard, off-shell line $i$ as an operator in the space of the involved
quantum numbers, and ${\mathbf R}_s$ encodes the remaining structure
we intend to factor from the hard process amplitude, {\it i.e.}
${\mathbf R}_s$ contains all couplings, numerator structures and gauge
structures, except the numerators of the hard propagators ${\mathbf
  P}_i$ attaching to it.  This factorization (in \eqref{eq:offshell}) --- which
has not yet used any approximation --- can be diagrammatically represented
as \beq \raisebox{-32 pt}{\includegraphics[scale=0.45]{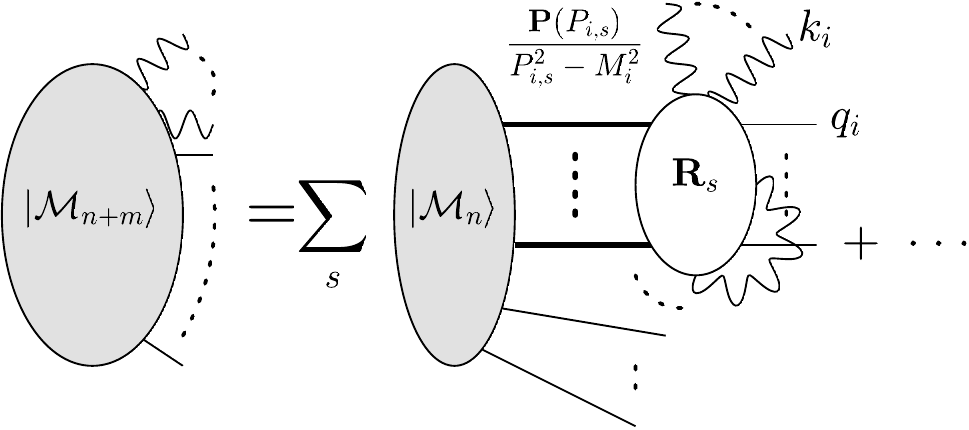}}
\label{eq:exchange-kinematics}
\eeq where wavy lines denote unresolved particles,
$P_{i,s}=q_i+K_{i,s}$ and the ellipse with ${\mathbf R}_{s}$ refers to topologies which
do not factor separately onto the $n$-parton amplitude.

Our aim is
to identify when --- in a very general setting --- this amplitude factors in
a systematically expandable way onto an {\it on-shell} hard amplitude
after isolating external sub-diagrams as above.

We also discuss how we can
construct bases for the space of chiral structures 
such that we can
express the abstract operators in a concrete fashion and iterate
virtual exchanges and emissions in the solution to an evolution
equation of the amplitude.

\subsection{Kinematics}
\label{sec:kinematics}

Before we address the more complicated electroweak case, let us
recall how soft factorization in QCD works. In this case we would consider
emissions and exchanges which contribute a total momentum $K_{i,s}$ to
an off-shell line $i$, carrying a total momentum $P_{i,s}= q_i + K_{i,s}$,
where $q_i$ is the on-shell external momentum. Thus
\begin{equation}
  \frac{{\cal M}(P_{i,s})}{P_{i,s}^2 - M_i^2} = \frac{{\cal M}(q_i + K_{i,s})}{2 q_{i}\cdot K_{i,s} + K_{i,s}^2} \ ,
  \label{eq:prop1}
\end{equation}
which follows from $q_i^2 = M_i^2$. In the uniform soft limit,
$K_{i,s}\to \lambda K_{i,s}$, we differentiate w.r.t. $\lambda$
to get the power expansion in $\lambda$ 
\begin{multline}
  \frac{{\cal M}(P_{i,s})}{P_{i,s}^2 - M_i^2} \to
  \frac{1}{\lambda}\frac{{\cal M}(q_i)}{2 q_{i}\cdot K_{i,s}}
  +\\
  \frac{2 q_{i}\cdot K_{i,s} ( K_{i,s}^\mu \partial_\mu {\cal M}(q_i)) - K_{i,s}^2 {\cal M}(q_i)}{4 (q_{i}\cdot K_{i,s})^2} + {\cal O}(\lambda) \ .
   \label{eq:prop2}
\end{multline}
Going back to \eqref{eq:prop1}, we note that we can alternatively
parametrize the hard momentum $q_i$ by a different hard direction
$p_i$ on the same mass-shell,
\begin{equation}
  q_i^\mu = p_i^\mu - K_{i,s}^\mu + \frac{K_{i,s}\cdot( K_{i,s}-2 p_i)}{2n\cdot (K_{i,s}- p_i)} n^\mu
\end{equation}
with a light-like reference vector $n$, constrained only by
$n\cdot (K_{i,s}- p_i)\ne 0$. The above expression essentially serves
to obtain $q_i^2=p_i^2=M_i^2$ while
\begin{equation}
  q_i + K_{i,s} = p_i + \text{(recoil paramter)} n \ ,
\end{equation}
allowing us to expand around a new, on-mass-shell, hard direction
$p_i$, irrespective of the mass-shell condition on $K_{i,s}$ or
precisely the way we define a 'soft' $K_{i,s}$. All that matters is
that the recoil parameter is small,
\begin{equation}
|2 p_{i}\cdot K_{i,s}  - K_{i,s}^2 | \ll |n\cdot(p_i - K_{i,s}) |
\end{equation}
such as to obtain an on-shell amplitude from expanding
\begin{equation}
  \frac{{\cal M}(P_{i,s})}{P_{i,s}^2 - M_i^2} = \frac{n\cdot(p_i - K_{i,s})}{n\cdot p_i}
  \frac{{\cal M}(p_i + (...)n)}{2 p_{i}\cdot K_{i,s}-K_{i,s}^2 } .
\end{equation}
In fact, for the soft gluon case above the new parametrization will
lead to a similar leading-power expansion as
$K_{i,s}\to \lambda K_{i,s}$, $\lambda\to 0$,
\begin{equation}
  \frac{{\cal M}(P_{i,s})}{P_{i,s}^2 - M_i^2} \to
  \frac{1}{\lambda}\frac{{\cal M}(p_i)}{2 p_{i}\cdot K_{i,s}}
  + {\cal O}(\lambda^0) \ ,
\end{equation}
with $q_i = p_i + {\cal O}(\lambda)$, however it will differ at
subleading power.

The aim of our work is thus to make the above procedure an exact
parametrization which we can use for a leading power expansion, subject
to an explicit Eikonal propagator, including a possible change of mass
shell between the emitter momentum before and after the emission,
$p_i^2\ne q_i^2$, and subject only to maintaining momentum
conservation of all momenta attached to the amplitude.

As seen, in general the parametrization of the kinematics is
complicated by the mass-shell conditions. We consider $q_i^2=m_i^2$
for the external particles,
while the particles propagating along the off-shell lines have masses
$M_i$ (as they may well have other flavors).  We note that these
masses refer to {\it physical, on-shell masses}, a choice which will
provide us with a factorization of physical, renormalized S-matrix
elements
\footnote{Effectively, the relation between kinematics and masses does
force us to use pole masses despite they might not be optimal. Keeping
track of renormalization factors and a controlled expansion we have,
however, all information at hand to convert to different mass
schemes.}.

On top of this, we need to allow for the possibility to
implement recoil to respect overall energy-momentum
conservation.

To understand the kinematic regions where the factorization
is applicable, we consider a frame where the off-shell
momentum, directly to the right of the big gray blob on the
the right-hand side in \eqref{eq:exchange-kinematics}, is
approximated by

\begin{equation}
  p_i = \left(\sqrt{E_{i}^2+M_i^2},\vec{0}_\perp, E_{i}\right)\;
\end{equation}
with $p_i^2=M_i^2$, while $P_{i,s}^2\ne M_i^2$. 

We then introduce a light-like momentum $n_{i,s}$ with a direction
which maximizes
$ p_i \cdot n_{i,s}$\;,

\begin{equation}
  n_{i,s} = \frac{n_{i,s}\cdot
    p_i}{E_{i}+\sqrt{E_{i}^2+M_i^2}}\left(1,\vec{0}_\perp,-1\right)\;.
\end{equation}
Any additional deviation from the four-momentum $p_i$ we
write in terms of 
\begin{equation}
  Q_{i,s} =
  \left(Q_{i,s}^{(+)}+Q_{i,s}^{(-)},\vec{Q}_{i,s}^{(\perp)},Q_{i,s}^{(+)}-Q_{i,s}^{(-)}\right) \ .
\end{equation}
With this in mind, we express the external momenta $q_i$ and the
$K_{i,s}$ in a covariant way as
\begin{eqnarray}
  \label{eq:mapping}
    K_{i,s}^\mu & = & \Lambda^\mu {}_\nu \left(Q_{i,s}^\nu + \delta_{i,s}\ n_{i,s}^\nu\right)\\\nonumber
    q_i^\mu &= & \Lambda^\mu {}_\nu \left(\alpha p_i^\nu + \frac{(1-\alpha^2)M_i^2 +
      p_i\cdot Q_{i,s}}{2 \alpha\ n_{i,s}\cdot p_i}n_{i,s}^\nu\right) - K_{i,s}^\mu\;,
\end{eqnarray}
where the parameter $\delta_{i,s}$ is determined such that $q_i^2=m_i^2$,
giving
\begin{eqnarray}
  \delta_{i,s} &=& \frac{(M_i^2\alpha^2 -m_i^2 + (1-2\alpha) p_i\cdot Q_{i,s} + Q_{i,s}^2)}{2 n_{i,s}\cdot(\alpha p_i - Q_i)}  \nonumber \\    
  &-&\frac{ n_{i,s}\cdot Q_{i,s} (M_i^2(1-\alpha^2)+p_i\cdot Q_{i,s})}
        {2\alpha\ n_{i,s}\cdot p_i\ n_{i,s}\cdot(\alpha p_i - Q_i)} \ ,
\end{eqnarray}
whereas the parameter $\alpha$, as well as the boost itself, 
relates to maintaining energy and momentum conservation, as 
discussed below.

For the momenta not involved in the exchange or emission (i.e., legs not in $h_s$),
setting $Q_{i,s}=0$ and $M_i^2=m_i^2$ in \eqref{eq:mapping} gives
\begin{equation}
  q_i^\mu=\Lambda^\mu {}_\nu \left(\alpha\ p_i^\nu +
  \frac{(1-\alpha^2)m_i^2}{2\alpha\ n_{i,s}\cdot p_i} n_{i,s}^\mu\right)\;.
\end{equation}

Note that, if emissions are involved, the
$K_{i,s}$ are some combination of emission and exchange momenta, which
satisfy $\sum_{i\in h_s} K_{i,s} = \sum_I k_i$ where the right hand
sum is over all emissions.  Thus the total outgoing momentum of our
process, $Q$, is
\begin{eqnarray}
 &&Q^\mu=\sum_i k_i^\mu + \sum_i q_i^\mu =\alpha \Lambda^\mu {}_\nu \Big(\\
  &&  \sum_{i} p_i^\nu + \frac{1-\alpha^2}{\alpha^2}
   \underbrace{\sum_i \frac{M_i^2}{2 n_{i,s}\cdot p_i} n_{i,s}^\nu}_{\equiv N_1^\nu}
  + \frac{1}{\alpha^2}\underbrace{\sum_{i\in h_s} \frac{p_i\cdot Q_{i,s}}{2 n_{i,s}\cdot p_i}n_{i,s}^\nu}_{\equiv N_2^\nu} \Big) \,.\nonumber 
\end{eqnarray}
In order to implement four-momentum conservation the Lorentz transformation and
scaling parameter $\alpha$ need to obey 
\begin{equation}
  (\Lambda^{-1})^\nu{}_\mu Q^\mu
  = \alpha\ Q^\nu + \frac{1-\alpha^2}{\alpha} N_1^\nu + \frac{1}{\alpha} N_2^\nu \ 
\end{equation}
which fixes $\alpha$ by requiring the same invariant mass
before and after the Lorentz transformation.

Demanding that $\alpha > 0$, and $\alpha\to 1$ if all masses
$M_i$, and the momenta $Q_{i,s}$ vanish, we find (cf. \eqref{eq:mapping}) a solution which in
general admits $\alpha= 1+{\cal O}(\lambda)$ as $p_{i,s}\cdot Q_{i,s}\to
\lambda\ p_{i,s}\cdot Q_{i,s}$, $\lambda \to 0$, irrespective of the
kinematic limit covered by this scaling. Note, however, that we have
not constrained the form of the Lorentz transformation (in particular it need
not be small), and in general only use
\begin{eqnarray}
  \label{eq:limitingkinematics}
  q_i^\mu &=& \Lambda^\mu {}_\nu \left( p_i^\nu - \tilde{K}_{i,s}^\nu + {\cal O}(\lambda)\right)
  \qquad \nonumber \\
  K_{i,s}^\mu &=& \Lambda^\mu {}_\nu \left( \tilde{K}_{i,s}^\nu
  +{\cal O}(\lambda)
  \right)\qquad \nonumber \\
  \tilde{K}_{i,s}^\nu &=& 
  \frac{M_i^2-m_i^2 + Q_{i,s}^2}{2(p_i\cdot n_{i,s}- n_{i,s}\cdot Q_{i,s})}n_{i,s}^\nu+Q_{i,s}^\nu
  \ .
\end{eqnarray}

Phase space factorization can be obtained systematically at leading
power in $\lambda$ for such kinematic mappings as shown in
\cite{Loschner:2021keu}.  We note that we can uniquely invert the
mapping and obtain an expression of $Q_{i,s}$ if we have fixed $p_i$
and $n_{i,s}$. This also means that we can use this definition also when
$K_{i,s}$ and $K_{j,s}$ are not independent, {\it e.g.}  for a
one-loop exchange in between two legs $i$ and $j$ we have $K_{i,s} = -
K_{j,s} = k$. Notice that we have chosen a backward direction
$n_{i,s}$ differently per hard momentum $p_i$.

The mapping in \eqref{eq:mapping} is designed, such that the denominator of
the off-shell propagators are directly given in terms of
\begin{equation}
  \label{eq:denom}
    (q_i+K_{i,s})^2 - M_i^2 = 2 p_i\cdot Q_{i,s} \ ,
\end{equation}
in analogy with \eqref{eq:prop2}.
Our expansion is thus in  
\begin{equation}
  \label{eq:counting}
  p_i\cdot Q_{i,s} \ll p_i\cdot n_{i,s} \equiv S_{i,s}\;
\end{equation}
and 
$\lambda\sim p_i\cdot Q_{i,s} /S_{i,s}$ is
our counting parameter which simultaneously enforces the above
hierarchy for all hard lines $i$.  It is important
to stress that we do not consider different $p_i$ for different
classes of diagrams $s$, while we may want to exploit different
parametrizations of unresolved momenta if needed.

\subsection{Factorization}

In order to
encode the quantum numbers $s$ of external particles (spin, color,
isospin, etc.) we introduce an operator corresponding to the on-shell
wave functions of the particles we consider,
\begin{equation}
  \langle s | \bar{\Psi}(q,m)|s'\rangle = \bar{\psi}_{s}(q,m)\delta_{ss'} \ ,
\end{equation}
\begin{equation}
  \label{eq:propcomplete}
  i \Psi(q,m)\bar{\Psi}(q,m) = \left. {\mathbf P}(q,m)\right|_{q^2=m^2} \ .
\end{equation}

In total, this allows us to write, at leading power,
\begin{widetext}
\begin{multline}
  \left(\prod_{i\notin h_s}\bar{\Psi}_{f_i}(q_i,m_i)\right)
  \left(\prod_{i\in h_s}\frac{{\mathbf P}_{f'_i}(q_{i}+K_{i,s},M_i)}{(q_{i}+K_{i,s})^2-M_{i}^2}\right)
   |{\cal M}_{\{f_i'\}_n} (\{P_i\}_n)\rangle = \\
   \left(\prod_{i\in h_s}\frac{\Psi_{f'_i}(\Lambda p_i,M_i)}{2 p_i\cdot Q_{i,s}}\right)\times \left(\prod_{i}\bar{\Psi}_{f'_i}(p_i,M_i)
   |{\cal M}_{\{f_i'\}_n} (\{p_i\}_n)\rangle\right) + {\cal O}(\lambda^{-\# h_s+1}) \ ,
\end{multline}
\end{widetext}
where $\#h_s$ is the number of hard off-shell legs interacting with
unresolved partons.  Here we have used that the amplitude contracted
with external wave functions is a Lorent invariant. In the light of
the kinematics discussion above, we then find that we can factor the
amplitude at leading power in $\lambda$ as
\begin{multline}
  |\tilde{\cal M}(\{q_i\}_n;\{k_i\}_m)\rangle \simeq\\\sum_s
  \mathbf{S}_{s}(\{q\}_{i\in h_s},\{k_i\}_m)   |\tilde{\cal M}(\{p_i\}_n)\rangle\;,
\end{multline}
in terms of the on-shell amplitude with $n$ external hard lines,
$|\tilde{\cal M}\rangle = \prod_i \bar{\Psi}_i|{\cal M}\rangle$, carrying
momenta $\{p_i\}_n$ (we have suppressed the flavor labels for the
sake of readability). The factored contribution is given by the
operator 
\begin{eqnarray}
  &&\mathbf{S}_{s;\{f_i\}_n,\{g_i\}_m}^{\{f'_i\}_n}(\{q\}_{i\in h_s},\{k_i\}_m) =
  i \left(\prod_{j\in h_s}\bar{\Psi}_{f_j}(q_i,m_i)\right)\nonumber \\
  &&\times {\mathbf R}_{s;\{f_i\}_n,\{g_i\}_m}^{\{f'_i\}_n}(\{q\}_{i\in h_s},\{k_i\}_m)
\left(\prod_{j\in h_s}\frac{{\Psi}_{f'_j}(\Lambda p_j,M_j)}{2p_j\cdot Q_{j,s}}\right) \ ,\nonumber\\
\end{eqnarray}
which is to be understood by expressing either $q_i$ or $p_i$ as a
function of $Q_{i,s}$, $n_{i,s}$ and $p_i$ or $q_i$ respectively. This
is possible because the amplitude contracted with external wave
functions, $|\tilde{\cal M}\rangle$, is Lorentz invariant, hence a
function of momentum invariants, and can thus be Taylor-expanded around
$p_i\cdot Q_{i,s}\to \lambda p_i\cdot Q_{i,s}$ in the limit
$\lambda\to 0$. Notice that also $q_i$ and the mapped momentum of the
exchanges and emissions are proportional to the Lorentz transform
which will therefore drop out of the final expression of our effective
matrix elements.  In essence we achieve factorization by evaluating
different parts of the amplitude in different frames. Furthermore, the
evaluation of the amplitude may be performed in different frames for
different sets $s$, and the phase space integration may be performed
in a third, as long as this difference is not contributing at leading
power.

Further simplifications can only occur if we consider stronger
constraints on the kinematic limits, though our factorization in this
general case serves as a starting point for a formula which
interpolates in-between different limits. In terms of our scaled
momentum $Q_{i,s}$, the requirement of \eqref{eq:counting} encompasses
kinematic configurations, which are essentially limited by several
different regions. 
They become apparent when one considers, in a specific frame for $p_i$
and $n_{i,s}$, the forward (along $p_i$) and backward (along
$n_{i,s}$) components $Q_{i,s}^{(\pm)}$ and the hard leg's
\textit{kinetic} energy $E_{i}$ and mass $M_i$, such that
$E_{i}^2+M_i^2\sim S_{i,s}$.

\begin{figure}
  \includegraphics[scale=0.4]{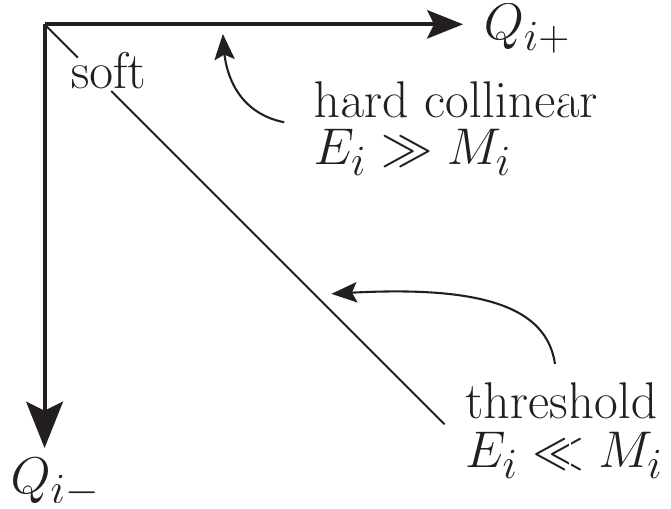}
  \caption{\label{fig:Q-plot}Illustration of the various regions of validity of our
    parametrization. As seen from, \eqref{eq:pi.Qi} the condition
    $p_i \cdot Q_{i,s} \ll p_i\cdot n_{i,s}$ is fulfilled if either both $Q_{i+}$ and
    $Q_{i-}$ are small (the genuinely soft region)
    or if $E_i\gg M_i$ and $Q_{i-}\rightarrow 0$ (the hard collinear region)
    or if $E_i\ll M_i $ and $Q_{i-}\rightarrow -Q_{i+}$ (the threshold region).
  }
\end{figure}

Our expansion is valid if
\begin{eqnarray}
  \label{eq:pi.Qi}
  p_i\cdot Q_{i,s}
  &=&\sqrt{E_{i}^2+M_i^2}(Q_{i,s}^{(+)}+Q_{i,s}^{(-)}) 
  + E_{i}(Q_{i,s}^{(-)}-Q_{i,s}^{(+)}) \nonumber\\
  &\ll& p_{i,s}\cdot n_{i,s}= S_{i,s} \ .
\end{eqnarray}
The regions of validity contain a Glauber-type region in which $Q_{i,s}$
becomes purely transverse, along with a soft, a hard-collinear,
and a threshold region, as depicted in \figref{fig:Q-plot}.

 One boundary of the available phase space
is the hard (quasi-)collinear region in which $Q_{i,s}^{(-)}\ll
S_{i,s}$ but $Q_{i,s}^{(+)}$ is unconstrained and the hard leg is
highly energetic, $E_{i}\gg M_i$. Another limiting region is the
threshold region with $E_{i}\ll M_i$. The regions intersect in the
genuine 'soft' region where $Q_{i,s}^{(+)}\sim Q_{i,s}^{(-)}\ll
S_{i,s}/\sqrt{E_{i}^2+M_i^2}$ i.e., $Q_{i,s}^\mu$ is small compared to
the hard scales in all of its components.  The 'soft' region also
contains a Glauber-type region in which $Q_{i,s}$ becomes purely
transverse. In both cases, the exchange or emission momentum $K_{i,s}$
is then accounting for the change in mass-shell between $M_i$ and
$m_i$ in its respective forward and backward components. Note that
$K_{i,s}$ will in general not be soft in the usual sense, neither in
its three-momentum, nor in all of its components. In the quasi-soft
limit we will find a generalized Eikonal approximation in which
$Q_{i,s}$ is small in all of its components.  In this case we can
write
\begin{equation}
  q_i^\mu = \Lambda^\mu {}_\nu \left( p_i^\nu + \frac{m_i^2-M_i^2}{2p_i\cdot n_{i,s}}n_{i,s}^\nu + {\cal O}(\lambda)\right) \ .
\end{equation}
Also note that we do not rely on the hard line being highly energetic
or close to threshold.

\subsection{Self energy insertions, wave function renormalization, and cutting of unresolved lines}
\label{sec:self-energy}

Within our
factorization, one would also be tempted to consider the factorization
of self-energy insertions which appear as iterations of
\begin{equation}\nonumber
  \frac{{\mathbf P}_{f'_i}(q_{i}+K_{i,s},M'_i)}{(q_{i}+K_{i,s})^2-M_{i}^{'2}} {\mathbf \Sigma}_{f'_i f_i}(q_{i}+K_{i,s})\;
\end{equation}
on the leg with momentum $q_{i,s}+K_{i,s}$ in the figure in
\eqref{eq:exchange-kinematics}, or on a corresponding emission
process.  Note that the self energy ${\mathbf \Sigma}_{f'_i f_i}$ is
an operator in the space of quantum numbers, as well.  Multiple
insertions, however, are not separated in scale, but contribute
equally at leading power with the same propagator attached and
therefore show no hierarchy.  Contributing propagators from different
intermediate particles would appear to be suppressed if the masses of
the mixing particles are different, $(q_{i}+K_{i,s})^2-M_{i}^{'2} = 2
p_i\cdot Q_{i,s} + M_i^2-M_i^{'2}$, however propagators which resum
these effects develop poles at the mass shells of all particles
involved in the mixing. This necessarily leads us to consider resummed
propagators and a proper relation to physical masses. In fact, as
highlighted above, we parametrize the kinematics in terms of the
physical masses $m_i$ and $M_i$. In this case, the propagator
denominators, for a complex mass scheme
\cite{Denner:2006ic,Denner:2014zga}, are expressed in terms of the
renormalized (complex) mass parameters $M_{i,R}^2$ and renormalized
self energy contributions $\Sigma_{i,R}$, where the physical mass is a
solution to $M_i^2=M_{i,R}^2+\Sigma_{i,R}(M_i^2)$. Our mapping has the
virtue that
\begin{multline}
  \frac{1}{(q_i+K_{i,s})^2-M_{i,R}^2-\Sigma_{i,R}((q_i+K_{i,s})^2)} =\\
  \frac{1}{\lambda} \frac{1}{2p_i\cdot Q_{i,s}}\frac{1}{1-\Sigma'_{i,R}(M_i^2)} + {\cal O}(1)
\end{multline}
as $p_i\cdot Q_{i,s}\to \lambda p_i\cdot Q_{i,s}$, $\lambda\to 0$,
thus providing the proper wave function renormalization to the hard
amplitude we factor to, and the legs involved in the contributions we
do intend to factor from the amplitude. Residues of mixing propagators
can then be accounted for in the exchange or emission kernels together
with the elementary vertex. Beyond the leading order, the ${\mathbf S}$
operator will thus be provided with the relevant wave function
renormalization constants and as such is defined beyond the lowest
order. This holds for all internal lines we consider here (scalar,
fermion, vector), as well as for unstable particles when using a
complex mass scheme.

Another consequence is that the program of casting virtual
corrections into phase space type integrals to locally cancel infrared
enhancements from the real emission (as {\it e.g.} systematized in
\cite{Platzer:2020lbr}) thus faces an important modification: Instead
of an on-shell cut through the unresolved, `soft', exchanges we will
need to use a cutting rule
\begin{multline}
  \frac{1}{k^2-m^2-i m \Gamma\ \text{sign}(T\cdot k)} = \\
    \frac{1}{k^2-m^2+i m \Gamma} + 2i \frac{m\Gamma}{(k^2-m^2)^2+m^2\Gamma^2}\theta(T\cdot k) \ .
\end{multline}
This identity has a straightforward physical interpretation: while it
clearly yields the standard cut result for $\Gamma\ll m$, it instructs
us for the finite-width case to replace the cut with a Breit-Wigner
factor, and cuts through the decay products of the exchanged unstable
particle, noting that $2m\Gamma$ is the exchanged particle's decay
matrix element integrated over phase space. Unitarity as a building
block of parton branching and resummation algorithms thus appears in a
different form, though this poses no conceptual problem if one
treats subtraction terms for real and virtual corrections
separately, and performs a careful analysis of measurements
\cite{Platzer:2022jny}. The latter also will project onto
decays of the unstable physical bosons after their high energy
evolution.

\section{A complete flow picture}
\label{sec:flow}

In this section, we discuss (flow) versions of the bases for
color, chirality and isospin, to be used in \eqref{eq:amp}.

For the color structure, it is well known how to employ the Fierz
identity to decompose all color structure into flows, see for example
\cite{Platzer:2013fha,Platzer:2020lbr,DeAngelis:2020rvq,Frixione:2021yim,Forshaw:2021mtj}.
We remark, that while this paper focuses on flow representations,
one can of course also use other decompositions. In particular
for color structure, orthogonal bases can be used
\cite{Keppeler:2012ih,Alcock-Zeilinger:2016cva,Sjodahl:2018cca},
a context in which there has recently been significant development
\cite{Alcock-Zeilinger:2022hrk,Keppeler:2023msu}.

For isospin, we note that the chiral states allow us to work
directly with eigenstates of the isospin operator. At high
energies, in the unbroken phase, one could treat $SU(2)_L$
in terms of flows, as for any $SU(N)$.
However, as we want to treat the weak bosons as mass eigenstates,
we instead propose to simply work with explicit weak
isospin eigenstates. Hypercharge comes from an abelian $U(1)$
and therefore does not come with any flow representation.

Also the chiral structures from spin and
momenta can be decomposed into flows by employing the Fierz identity
on the spinors, a simplification which is built into the Feynman rules of
chirality flow. We note however, that for a complete reduction
in terms of \textit{tensor structures} we need more flows than
just the direct contraction between external spinors. As we will see, this
means that the tensor flow basis for chiral structures comes
with more terms than the flow basis for color.

Overall, Fierz rearrangements will thus allow us to obtain a basis of (tensor)
structures for quantities like
$\left(t^a\right)^{i_1} {}_{i_3}\left(t^a\right)^{i_2} {}_{i_4}\
\delta^{I_1}_{I_3} \delta^{I_2}_{I_4}\
\left(\overline{P_{c_1}}\gamma^\mu P_{c_3}\right)_{\beta \alpha }
\left(\overline{P_{c_2}}\gamma_\mu P_{c_4}\right)_{\delta \gamma}$ such
that we can express the amplitude in a basis of flows, $|\sigma\rangle$,
\begin{equation}
  |T\rangle = \sum_\sigma c_\sigma |\sigma\rangle \ ,
\end{equation}
\begin{equation}
  |{\cal M}\rangle =
  \sum_\sigma
{\cal M}_\sigma
  |\sigma\rangle \ ,
\end{equation}
with ${\cal M}_\sigma = G\ c_\sigma$ in our simple example from the
introduction, Sec.~\ref{sec:notation}, though more general and non-trivial
mixtures of kinematic dependence will multiply each flow vector in the
case of general amplitudes. 

\subsection{A flow basis for chiral structures}
The chiral nature of the electroweak interaction, and the relevance of
spin correlations, call for a flow concept which we will
introduce now: In analogy with performing resummation in color space
using a spanning set of color flows, we will prove that the
resummation evolution in Lorentz space can be described using
``chirality flows''.  We thus build on the chirality-flow formalism
\cite{Lifson:2020pai,Alnefjord:2020xqr,Lifson:2022ijv,Boman:2023afu,Lifson:2023wow}, which allows
the immediate translation of Feynman diagrams to spinor inner
products.

We therefore describe particles in terms of their chirality, and
expect a decomposition of the full amplitude (with both left- and
right helicity) to chirality to have been performed before the start
of the evolution. To be precise we want to choose a basis for the
amplitude vector written as a vector in our abstract formalism above,
$\langle \langle \{i_1,...\},\{I_1,...\},\{\alpha,...\} |{\cal
  M}\rangle$, where $|{\cal M}\rangle$ is the amplitude without the
external wave functions ({\it cf.} a color flow without assigned
external colors) and we will work out the action of the factored
diagrams using \eqref{eq:propcomplete}. In this way, we will gain full
analytic control of the Lorentz structure.

Denoting
a left-chiral fermion with momentum $p_i$ with $|i]=\raisebox{-7
    pt}{\includegraphics[scale=0.4]{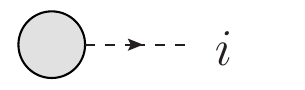}}$\hspace*{-0.4 cm}
  (or $[i|=\raisebox{-7
      pt}{\includegraphics[scale=0.4]{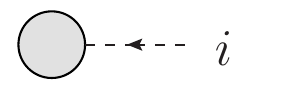}}$\hspace*{-0.4
      cm}) and a right-chiral fermion with $|j\rangle=\raisebox{-7
      pt}{\includegraphics[scale=0.4]{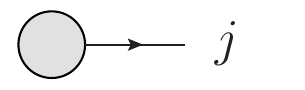}}$\hspace*{-0.4
      cm} (or $\langle j|=\raisebox{-7
      pt}{\includegraphics[scale=0.4]{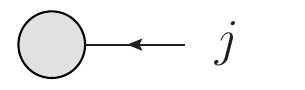}}$\hspace*{-0.4
      cm}) we want to consider the effect of (say) a photon exchange
    between two --- for now massless --- fermions. The effect of mass
    will be considered below.

Representing the Lorentz structure $p_\mu \sigma^\mu$ with a
``momentum dot'' $p_\mu \sigma^\mu=\raisebox{-3
  pt}{\includegraphics[scale=0.4]{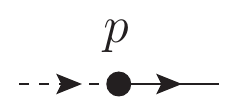}}$, and
similarly $p_\mu \bar{\sigma}^\mu=\raisebox{-3
  pt}{\includegraphics[scale=0.4]{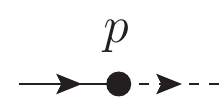}}$, we have, for
an exchange between two legs, the chiral structure (drawn in black on
top of a gray Feynman diagram) to the left below for two left-chiral
fermions and the structure to the right if $i$ is left-chiral, and $j$
is right-chiral \beq \raisebox{-20
  pt}{\includegraphics[scale=0.4]{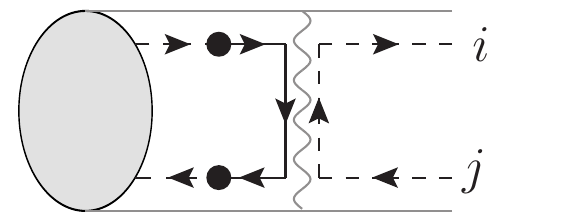}}
\raisebox{-20
  pt}{\includegraphics[scale=0.4]{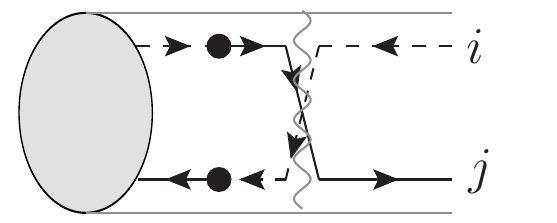}}
.
\label{eq:photon-exchange}
\eeq Here, to the left, the dashed line connecting the outgoing
particles $i$ and $j$ is the graphical representation of the spinor
inner product $[j\,i]$. After the exchange, the particles $i$ and $j$
are thus connected by a ``chirality flow''.  The momentum dots connect
somewhere within the blob and (naively) complicates the chirality
structure of the rest of the diagram. However, as we will show below,
a complete set of chirality-flow structures
connecting the external spinors can be given by considering the
contractions
\beq
\includegraphics[scale=0.4]{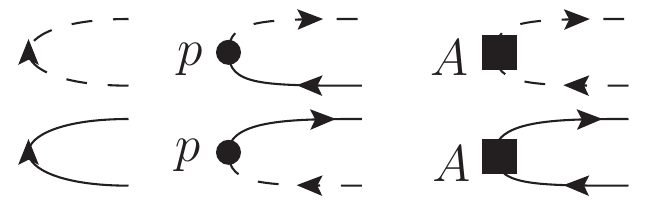}
\label{eq:flows}
\eeq
for some four-vector $p$ contracted with $\sigma/\sigmabar$,
and some antisymmetric rank two tensor
$A_{\mu \nu}$ contracted with
$\frac{1}{2}\left(\sigma^\mu \sigmabar^\nu -\sigma^\nu \sigmabar^\mu\right)$, and
for connections between all pairs of external particles. Before the exchange,
the particles $i$ and $j$ to the left in \eqref{eq:photon-exchange} were thus contracted to some
(other) external particles via these structures.

After the exchange, the chirality flows (of the type in
\eqref{eq:flows}) to which $i$ and $j$ were contracted, will be
connected to each other via the double momentum-dot structure in the
left diagram.  This gives rise to structures with up to 2+2+2 Lorentz
index contractions (in case $i$ and $j$ connected to two different
chirality-flow structures of type $\raisebox{-3
  pt}{\includegraphics[scale=0.4]{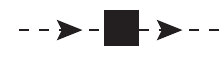}}$).

In case the external particles have
opposite chirality, we will have a chirality flow of the
type to the right in \eqref{eq:photon-exchange},
giving rise to two momentum dot structures of up to
2+1 Lorentz indices (if $i$ and $j$ were originally chirality-flow
connected to say $i'$ and $j'$ respectively via
$\raisebox{-3 pt}{\includegraphics[scale=0.4]{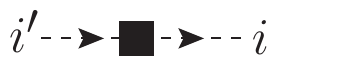}}$\hspace*{-3mm},
$\raisebox{-3 pt}{\includegraphics[scale=0.4]{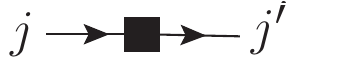}}$\hspace*{-3mm},
or connected to each other via
$\raisebox{-3 pt}{\includegraphics[scale=0.4]{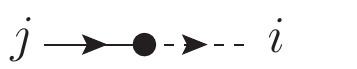}}$\hspace*{-5mm}
).

We will now argue that in both cases, the structures can be simplified
back to the cases in \eqref{eq:flows}.  We also schematically derive
the decomposition of structures with up to six momentum dots which can
appear in intermediate steps.

First, we note that the usage of
$\la i\,j\ra, [i\,j]$;
$p_\mu [ i\,| \sigma^\mu |j \ra, p_ \mu\la i\,|\sigmabar^\mu|j]$ and
$A_{\mu\nu} \la i\,| \sigmabar^{[\mu} \sigma^{\nu]}|j \ra$,  $A_{\mu\nu} \la i\,|\sigma^{[\mu} \sigmabar^{\nu]}|j]$
  for some antisymmetric tensor $A^{\mu\nu}$,
  is equivalent to the decomposition of products of $\gamma$-matrices into
  $1$ and $\gamma^5$ (via the decomposition into left and right chiral states);
  $\gamma^\mu,\;\gamma^5\gamma^\mu$ and
  $[\gamma^\mu, \gamma^\nu]$ respectively.

  In principle this is in itself a proof that these are the structures to
  anticipate. Nevertheless we will explicitly prove that the action
  of gauge boson exchange, starting from any of these structures, will
  result in linear combinations of the same spanning structures.
  That this holds for exchange of fermions and scalars then follows
  trivially.
  We start with considering the simplest case when two momentum dots,
  for example coming from the fermion propagators (slashed momenta)
  arising by photon exchange,
  are attached to the same chirality-flow line.
  Using the chirality-flow Feynman rules from \cite{Lifson:2020pai},
  we obtain
  \beq
  \raisebox{-30pt}{  \includegraphics[scale=0.4]{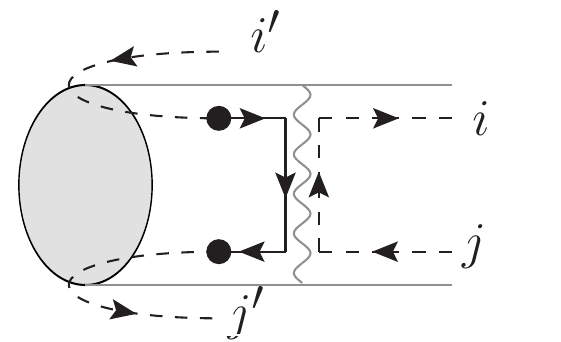} }\hspace*{-2mm},
  \eeq
  i.e. the partners $i'$ and $j'$ (originally connected to $i$ and $j$
  respectively) become chirality-flow connected, whereas $i$ and $j$ instead becomes connected
  to each other. To decompose the two momentum dots into our basis, we use the
  identity
  \beq
  \sigma^\mu \sigmabar^\nu=g^{\mu\nu}+\frac{1}{2} (\sigma^\mu \sigmabar^\nu-\sigma^\nu \sigmabar^\mu)\;.
  \eeq
  Contracting with external momenta $p_1$ and $p_2$, and sandwiching between external spinors
  this gives
  \begin{eqnarray}
  \underbrace{[ i\,|p_{1\mu} p_{2\nu}\sigma^\mu \sigmabar^\nu|j]}_{\equiv\raisebox{-0.2cm}{\includegraphics[scale=0.4]{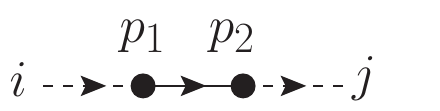}}}
  &=&p_{1} \cdot p_{2}\hspace*{-0.5cm}\underbrace{[ i\,j]}_{\quad\equiv\raisebox{-0.2cm}{\includegraphics[scale=0.4]{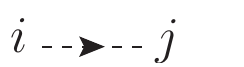}}} \\ \nonumber
  &+&\hspace*{+0.4cm}\underbrace{[ i\,| p_{1\mu} p_{2\nu}\frac{1}{2} (\sigma^\mu \sigmabar^\nu-\sigma^\nu \sigmabar^\mu)  |j]}_{\equiv\raisebox{-0.4cm}{\includegraphics[scale=0.4]{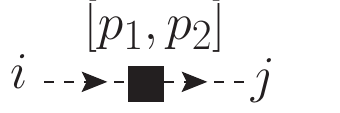}}}\;
  \label{eq:dot-dot-decomp}
  \end{eqnarray}
  (applied to $i'$ and $j'$). We remark that it is the antisymmetric part of
  $p_{1\mu}p_{2\nu}$, $\frac{1}{2}(p_{1\mu}p_{2\nu}-p_{1\nu}p_{2\mu})$, that survives
  the contraction in the second term; more generally,
  we find an antisymmetric rank-2 tensor contracted with
  $\frac{1}{2} (\sigma^\mu \sigmabar^\nu-\sigma^\nu \sigmabar^\mu)$.
  
  Exploring the effect on the basis vector we obtain
  \begin{eqnarray}
    & &\raisebox{-22 pt}{\includegraphics[scale=0.4]{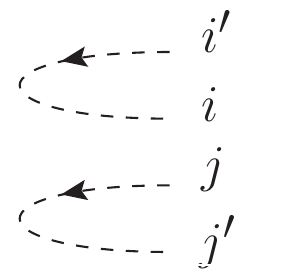}}
    \rightarrow
    \raisebox{-24 pt}{\includegraphics[scale=0.4]{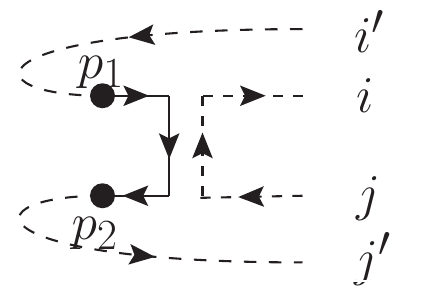}}\\ \nonumber
    &=&\;
    p_1\cdot p_2 \raisebox{-22 pt}{\includegraphics[scale=0.4]{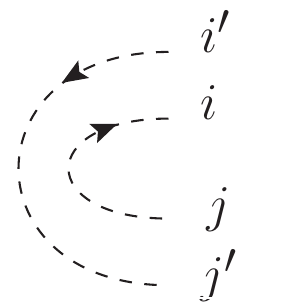}}
    +
    A_{12}\raisebox{-22 pt}{\includegraphics[scale=0.4]{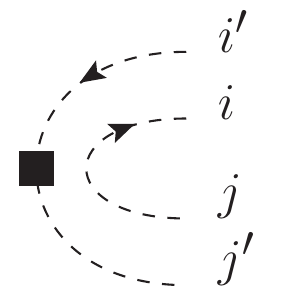}}
  \end{eqnarray}
  with
  \begin{equation}
    (A_{12})_{\mu\nu}=\frac{1}{2}(p_{1\mu}p_{2\nu}-p_{1\nu}p_{2\mu})\;,
  \end{equation}
  in the typical case that $i$ and $j$ are not chirality-flow connected to each other.
  If they are, we find for example
  \begin{eqnarray}
    \raisebox{-10 pt}{\includegraphics[scale=0.4]{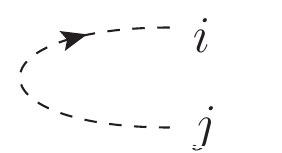}}
    &\rightarrow& 
    \raisebox{-17 pt}{\includegraphics[scale=0.4]{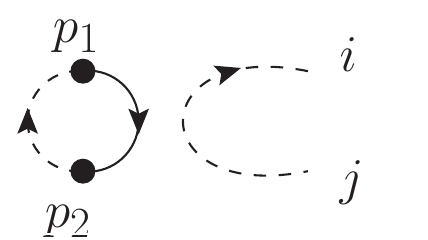}}\;\nonumber \\
    &=& 2  p_1\cdot p_2 \raisebox{-10 pt}{\includegraphics[scale=0.4]{1-delta-initial-flow}}\;,
  \end{eqnarray}
  i.e. we get back the chirality flow that we start with.     
  
  When describing the effect of spin-1 exchange, we also encounter
  structures with three momentum
  dots, for example while exchanging a photon between two fermions ($i$ and $j$)
  which are chirality-flow connected with partners ($i'$ and $j'$), in one case directly
  and in the other with a single momentum-dot in-between,
  
  \begin{eqnarray}
    \raisebox{-32 pt}{\includegraphics[scale=0.4]{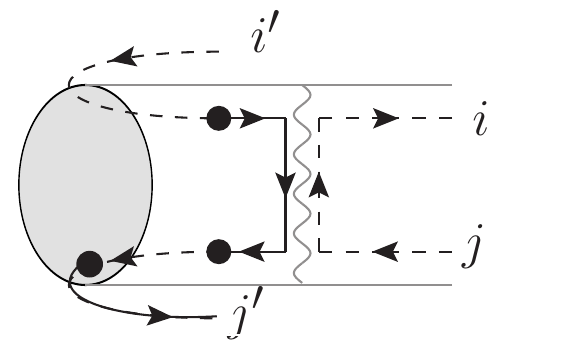}} .
    \label{eq:3dot-orig}
  \end{eqnarray}
  The line with three bullets represents the contraction
  $p_{1 \mu} p_{2 \nu} p_{3 \rho} [i'\,| \sigma^\mu \sigmabar^\nu \sigma^\rho|j' \ra$
    for the four-vectors $p_1,\,p_2$ and $p_3$ associated with the momentum-dots.
    To decompose this structure, we contract     
    $\sigma^\mu \sigmabar^\nu \sigma^\rho$
    with $g_{\mu\nu}, g_{\mu \rho},  g_{\nu\rho}$ and
    $\epsilon_{\mu\nu\rho\alpha}$, 
    giving a system of four equations and four unknowns with solution
    \begin{eqnarray}
      \sigma^\mu \sigmabar^\nu \sigma^\rho
      =g^{\mu\nu} \sigma^\rho-g^{\mu\rho} \sigma^\nu+g^{\nu\rho} \sigma^\mu
      +i \epsilon^{\mu \nu \rho}_{\;\;\;\;\;\alpha}  \sigma^\alpha\;,
      \label{eq:3sigma-decomp}
    \end{eqnarray}
    corresponding to
    \begin{eqnarray}
      &p_{1 \mu}& p_{2 \nu} p_{3 \rho} [k\,| \sigma^\mu \sigmabar^\nu \sigma^\rho|j \ra  \\
        &=&
        p_1\cdot p_2 \, p_{3 \rho} [k\,|\sigma^\rho|j \ra
          - p_1\cdot p_3\, p_{2 \rho} [k\,|\sigma^\rho|j \ra \nonumber \\
            &+& p_2\cdot p_3 \,p_{1 \rho} [k\,|\sigma^\rho|j \ra
              +i \underbrace{p_{1 \mu} p_{2 \nu} p_{3 \rho} \epsilon^{\mu \nu \rho}_{\;\;\;\;\;\alpha}}_{p_{123\,\alpha}}  [k\,|\sigma^\alpha|j \ra\;, \nonumber 
    \end{eqnarray}
    or in the momentum-dot notation (for general $[k|$ and $|j\rangle$)
    \begin{eqnarray}
      &&\raisebox{-3 pt}{\includegraphics[scale=0.4]{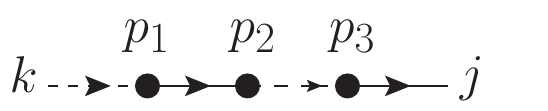}}  \nonumber\\
      &\quad& \quad = p_1\cdot p_2 \; \raisebox{-3 pt}{\includegraphics[scale=0.4]{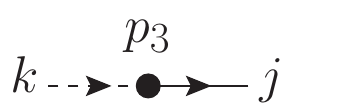}}\hspace*{-4mm} 
      - p_1 \cdot p_3 \; \raisebox{-3 pt}{\includegraphics[scale=0.4]{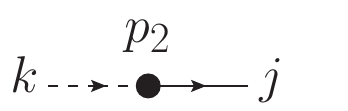}}\hspace*{-4mm}
      \nonumber \\
      &\quad& \quad +p_2 \cdot p_3\; \raisebox{-3 pt}{\includegraphics[scale=0.4]{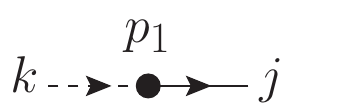}}\hspace*{-4mm}
      +i  \; \raisebox{-3 pt}{\includegraphics[scale=0.4]{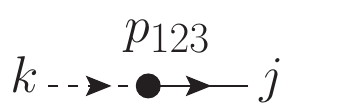}} \nonumber \\
      &\quad& \quad =\; \raisebox{-3 pt}{\includegraphics[scale=0.4]{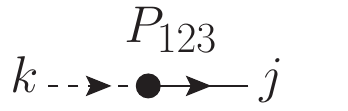}}
      \label{eq:3-dot-decomp}
    \end{eqnarray}
    for
    \beq
     P_{123}\equiv p_1 \cdot p_2\, p_3- p_1 \cdot p_3 \,p_2 +p_2 \cdot p_3 \,p_1 +i p_{123}\;.
    \eeq
    This means that we retrieve the structure of a single momentum-dot for
    some 'momentum' $P_{123}$.
   
    For the action of a momentum-dot on our basic building blocks, the basis vectors,
    \eqref{eq:flows} in the main text, we find by antisymmetrizing
    \begin{eqnarray}
      &&\raisebox{-3 pt}{\includegraphics[scale=0.4]{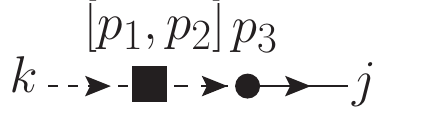}} \hspace*{-4mm}\nonumber \\
      &&=
      - p_1 \cdot p_3 \; \raisebox{-3 pt}{\includegraphics[scale=0.4]{3-mom-dot-RHS2}}\hspace*{-4mm}
      +p_2 \cdot p_3\; \raisebox{-3 pt}{\includegraphics[scale=0.4]{3-mom-dot-RHS3}}\hspace*{-4mm}
      \nonumber \\
    &&+i  \; \raisebox{-3 pt}{\includegraphics[scale=0.4]{3-mom-dot-RHS4}}.                 
    \label{eq:box-dot-decomp}
    \end{eqnarray}
    In the general case, the contraction to the left is with a general 
    antisymmetric rank-2 tensor $(A_{12})_{\mu_1 \mu_2}$.
    In this case, defining
    \beq
    P_{[[1,2],3]\mu}=
    2(A_{12})_{\mu \nu}p_3^{\nu}+i (A_{12})_{\mu_1 \mu_2} p_{3 \mu_3} {\epsilon^{\mu_1 \mu_2 \mu_3}}_{\mu}
    \eeq
    the decomposition reads
    \beq
    \raisebox{-3 pt}{\includegraphics[scale=0.4]{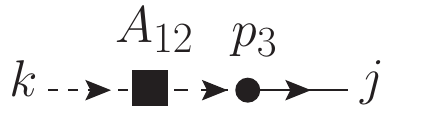}} \hspace*{-4mm}=
    \; \raisebox{-3 pt}{\includegraphics[scale=0.4]{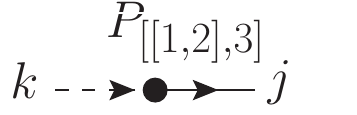}}.        
    \label{eq:box-dot-decompv2}
    \eeq
    The next structures to consider are those with a total of four contractions,
    either of form $\raisebox{-3 pt}{\includegraphics[scale=0.4]{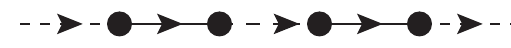}}$
    (where the middle two dots come from a fermion propagator, and the outermost
    dots from the initial chirality flows containing one momentum dot each),
    of form  $\raisebox{-3 pt}{\includegraphics[scale=0.4]{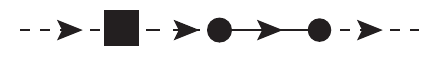}}$
    (from one initial chirality flow of type
    $\raisebox{-3 pt}{\includegraphics[scale=0.4]{box-no-label}}$, one or two
    momentum dots from a propagator, and (if one) the other from the initial chirality
    flow of the other involved fermion), or of form
    $\raisebox{-3 pt}{\includegraphics[scale=0.4]{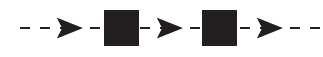}}$ (from an
    initial state of two boxes contracted with the mass term in the propagator).
             
    For this decomposition, we first explore the antisymmetric part
    $\raisebox{-3 pt}{\includegraphics[scale=0.4]{box-box-no-label}}$.
    Exploiting symmetries and fixing constants, this can be decomposed into 
    \begin{eqnarray}
      && \raisebox{-4 pt}{\includegraphics[scale=0.4]{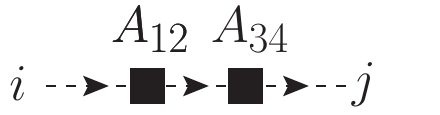}}\hspace*{-0.4cm} \nonumber \\
      &\quad& \quad =
      \left[-2(A_{12})_{\mu_1 \mu_2}(A_{34})^{\mu_1 \mu_2} \right. \nonumber \\
       &\quad& \quad + \left. i \epsilon^{\mu_1 \mu_2 \mu_3 \mu_4} (A_{12})_{\mu_1 \mu_2}(A_{34})_{\mu_3 \mu_4}\right] 
      \raisebox{-3 pt}{\includegraphics[scale=0.4]{2-mom-dot-RHS1}}\hspace*{-0.4cm}\nonumber \\
      &\quad& \quad+\raisebox{-4 pt}{\includegraphics[scale=0.4]{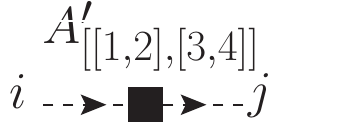}}
      \label{eq:box-box-decomp}
    \end{eqnarray}
    where
    \beq
    A'_{[[1,2],[3,4]]\,\mu \nu}=2\left( (A_{12})_{\mu \eta}{(A_{34})^{\eta}}_{\nu} -(A_{12})_{\nu \eta}{(A_{34})^{\eta}}_{\mu}\right)
    \label{eq:box-box-decomp-const}
    \eeq
    is made manifestly antisymmetric since only the antisymmetric part survives the
    contraction.
    The result in \eqref{eq:box-box-decomp}, along with \eqref{eq:dot-dot-decomp}
    and \label{eq:box-dot-decomp} (and versions where dotted and undotted lines
    switch role) completes the set of basic calculation rules needed to
    decompose any number of momentum dots  and boxes in a sequence.
    
    For example, using \eqref{eq:dot-dot-decomp} along with   
    \eqref{eq:box-box-decomp}
    the decompositions of
    $\raisebox{-3 pt}{\includegraphics[scale=0.4]{box-dot-dot-no-label}}$
    as well as 
    $\raisebox{-3 pt}{\includegraphics[scale=0.4]{4-mom-dot-no-label}}$             
    are straightforwardly obtained,
    \begin{eqnarray}
      &&\raisebox{-4 pt}{\includegraphics[scale=0.4]{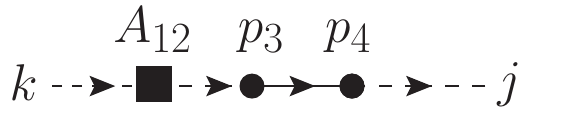}}\hspace*{-4mm}
      =p_3\cdot p_4  \raisebox{-4 pt}{{\includegraphics[scale=0.4]{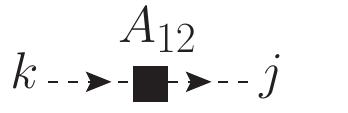}}}\hspace*{-4mm}\nonumber \\
      &&+\raisebox{-4 pt}{\includegraphics[scale=0.4]{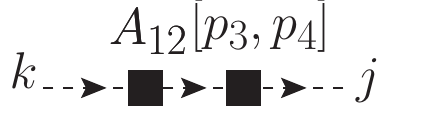}}\nonumber \;, \\
      &&\raisebox{-4 pt}{\includegraphics[scale=0.4]{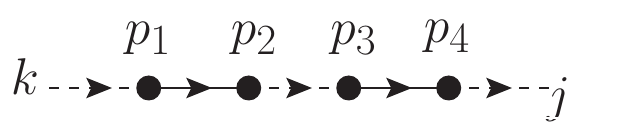}}\hspace*{-4mm}\nonumber \\
      &&=
      p_1 \cdot p_2\,p_3\cdot p_4 \raisebox{-3 pt}{\includegraphics[scale=0.4]{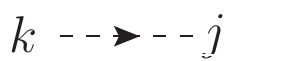}}\hspace*{-4mm}
      + p_3\cdot p_4 \raisebox{-4 pt}{\includegraphics[scale=0.4]{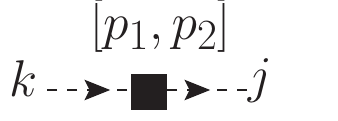}}\hspace*{-4mm}\nonumber \\
      &&+ p_1 \cdot p_2 \raisebox{-4 pt}{\includegraphics[scale=0.4]{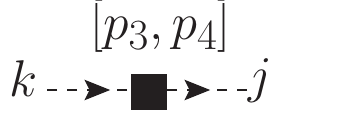}}\hspace*{-4mm}
      +\raisebox{-4 pt}{\includegraphics[scale=0.4]{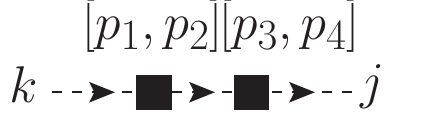}}\nonumber
      \label{eq:4dot-and--box-dot-dot}
    \end{eqnarray}
    where the double box structure may be expanded out using
    eqs.~(\ref{eq:box-box-decomp}) and (\ref{eq:box-box-decomp-const}).

    Similarly structures with in total five index contractions
    ($\raisebox{-3 pt}{\includegraphics[scale=0.4]{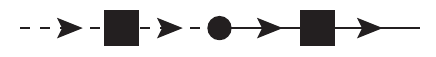}}$,
    $\raisebox{-3 pt}{\includegraphics[scale=0.4]{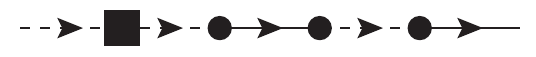}}$,
    as well as versions with dotted and undotted lines interchanged, and versions with
    arrows swapped)
    and 6 contractions
    (only $\raisebox{-3 pt}{\includegraphics[scale=0.4]{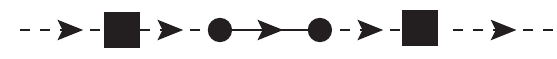}}$
    and a versions with dotted and undotted lines interchanged) can be obtained.
    
    In this way, the chirality-flow state obtained after several exchanges can
    iteratively be built up, and we obtain a chirality-flow decomposition analogous
    to the color-flow decomposition, but with the difference that there are three
    types of ``flows'' connecting partons, and that particles and anti-particles
    enter on equal footing.
    
    To illustrate this we  consider an example of intermediate complexity
    \beq
    \raisebox{-22 pt}{\includegraphics[scale=0.4]{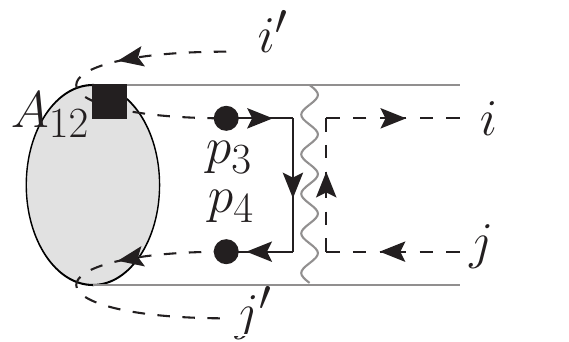}}\;,
    \eeq
    which, at the level of the basis vectors has the effect,
    \begin{eqnarray}
      &&\raisebox{-22 pt}{\includegraphics[scale=0.4]{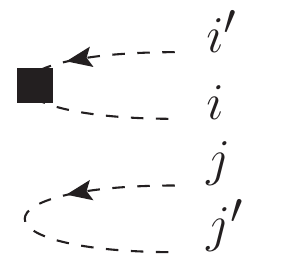}}\rightarrow \\
      && \left( -2(A_{12})_{\mu_1 \mu_2}(A_{34})^{\mu_1 \mu_2} 
      + i \epsilon^{\mu_1 \mu_2 \mu_3 \mu_4} (A_{12})_{\mu_1 \mu_2}(A_{34})_{\mu_3 \mu_4} \right)\nonumber \\
      &&\raisebox{-22 pt}{\includegraphics[scale=0.4]{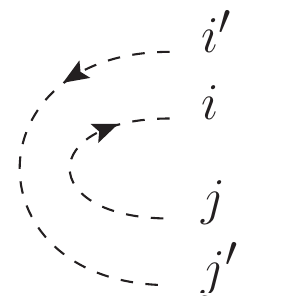}}\;
      +A'\raisebox{-22 pt}{\includegraphics[scale=0.4]{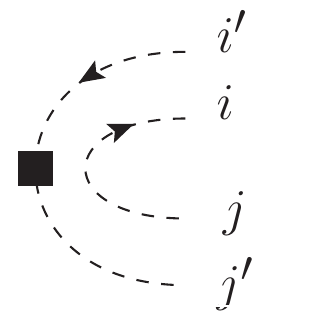}}\;
    \end{eqnarray}
    with
    \beq
    A_{34\,\mu\nu}=\frac{1}{2}(p_{3\mu}p_{4\nu}-p_{3\nu}p_{4\mu})
    \eeq
    and
    \beq
    \quad A'_{\mu \nu}=p_3\cdot p_4 A_{12 \mu \nu}+2\left( (A_{12})_{\mu \eta}{(A_{34})^{\eta}}_{\nu} -(A_{12})_{\nu \eta}{(A_{34})^{\eta}}_{\mu}\right)\nonumber
    \eeq
    as seen by expanding out 
    eqs.~(\ref{eq:box-box-decomp}) and (\ref{eq:box-box-decomp-const}) in
    \eqref{eq:4dot-and--box-dot-dot}.

In the light of the above description, combined with the chirality-flow
Standard Model Feynman rules \cite{Alnefjord:2020xqr},
it is clear that scalar
exchange, involving no chirality-flow line, does not change the
chirality flow (but alters the involved momenta).
Fermion exchange adds one (from the slashed momentum in the
propagator) or zero (from a potential mass term) momentum dots to
existing chirality-flow structures \cite{Alnefjord:2020xqr}.
For the mass term, expressed as a Kronecker delta in spinor indices, the line type
of the fermion line is left unchanged, but the chirality-flow
structure will change since (for example) chirality-flow lines which
are not originally connected may become connected.  External massive
fermions have to be decomposed into left- and right-chiral states,
as for example in \cite{Alnefjord:2020xqr}.

For the non-abelian vertices, we recall that they may be decomposed
into momentum-dot structures \cite{Lifson:2020pai,Alnefjord:2020xqr},
and therefore do not add to the number of possible chirality-flow
structures. (If a $W^\pm$, instead of a photon, is exchanged, the
chiral structure is rather simplified.)  For external gauge bosons, we
note that positive and negative helicity spin-1 particles appear as
one dotted and one undotted line with opposite directions, whereas the
longitudinal polarization of a massive vector boson can be expressed
in terms of a momentum dot \cite{Alnefjord:2020xqr}.  External gauge
bosons do, however, somewhat complicate the description, since they
may add a dependence on a reference gauge vector (which is unphysical
in the massless case, and related to the direction in which spin is
measured in the massive).

The conclusion is that none of the above pose a problem in
principle. Decomposing the original amplitude into the chirality-flow
objects in \eqref{eq:flows}, it is therefore possible to resum the
effect of soft interactions, much as the resummation is done in color
space using soft anomalous dimension matrices, with the basis vectors
being the structures in \eqref{eq:flows}, and the coefficients being
the vectors and tensors assigned to the momentum dots and the antisymmetric rank
two tensors.

In conclusion we thus find that the flow basis in \eqref{eq:flows} 
is applicable to resummation of all chiral structures
following after exchange of any known particle. In this sense,
this is the analogue of the color-flow basis.

\section{Conclusion and Outlook}
\label{sec:conclusion}

In this paper we have laid out the basis for performing amplitude
evolution within the electroweak Standard Model in order to account
for infrared enhanced contributions in a way similar to the soft gluon
resummation program in QCD. To achieve this, we build on the
chirality-flow formalism for treating the spin structure.  Kinematic
expansions are performed around the physical mass shells of particles
carrying a hard momentum, and include quasi-soft as well as other
enhanced kinematic regions. The resulting factorization of physical,
renormalized S-matrix elements accounts for color, isospin and spin
correlations, as well as the proper wave function renormalization
constants. Results of our formalism, together with suitable mappings
which implement energy-momentum conservation, can directly be
implemented in the \texttt{CVolver} evolution library
\cite{Platzer:2013fha,DeAngelis:2020rvq} and will serve as a basis to
design parton branching algorithms which include electroweak effects
beyond the quasi-collinear limit.

\section*{Acknowledgments}
We are thankful to Maximilian L\"oschner
for fruitful discussions (in particular on the complex mass scheme)
and for a very careful reading of the manuscript.
We are also thankful to Ines Ruffa for useful discussions.
This work was supported
by the Swedish Research Council (contract number 2016-05996, as well
as the European Union's Horizon 2020 research and innovation programme
(grant agreement No 668679). This work has also been supported in part
by the European Union’s Horizon 2020 research and innovation programme
as part of the Marie Skłodowska-Curie Innovative Training Network
MCnetITN3 (grant agreement no. 722104), and in part by the COST
actions CA16201 ``PARTICLEFACE'' and CA16108 ``VBSCAN''. We are
grateful to the Erwin Schr\"odinger Institute Vienna for hospitality
and support while significant parts of this work have been achieved
within the Research in Teams programmes ``Amplitude
Level Evolution II: Cracking down on colour bases.'' (RIT0521).

\bibliography{electroweak}

\newpage

\onecolumngrid
\appendix

\section*{Appendix}

\section{Propagators and external wave functions}

An important ingredient to our factorization formula is to
demonstrate, subject to the kinematic parametrization above, that
\begin{equation}
  \sum_{n=0}^\infty
  \left(\frac{{\mathbf P}(q_{i}+K_{i,s},M_i)}{(q_i+K_{i,s})^2-\tilde{M}_{R,i}^2}{\mathbf \Sigma}(q_{i}+K_{i,s})\right)^n
  \frac{{\mathbf P}(q_{i}+K_{i,s},M_i)}{(q_i+K_{i,s})^2-\tilde{M}_{R,i}^2} =
  \frac{1}{2p_i\cdot Q_{i,s}}
  \frac{\Psi(\Lambda p_i,M_i)\bar{\Psi}(\Lambda p_i,M_i)}{1-\Sigma'(M_i^2)} + {\cal O}(\lambda) \ ,
\end{equation}
where the derivative of the (physical part of the) self-energy
$\Sigma(p^2)$ (or, accordingly the transverse self-energy at vanishing
$k^2$ for a massless boson) provides the proper wave function
renormalization for the amplitude we factor to. To illustrate this let
us first consider Goldstone bosons in an $R_\xi$ gauge, with a free
propagator $i/(k^2-\xi\tilde{M}_{R,i}^2)$, where $\tilde{M}_{R,i}^2 =
M_{R,i}^2+i M_{R,i} \Gamma_{R,i}$ in a complex mass scheme
\cite{Denner:2006ic,Denner:2014zga}, and the introduction of
$\Gamma_{R,i}$ needs to be added back as additional insertions of
two-point functions. This does not provide any change to our main
argument. The propagators of the physical scalar can be obtained by
putting $\xi=1$. If the scalar has a one-particle irreducible
two-point function $-i\Sigma_S(k^2)$, the resummed propagator is
\begin{equation}
  \frac{1}{(q_i+K_{i,s})^2-\xi \tilde{M}_{R,i}^2 - \Sigma_S((q_i+K_{i,s})^2)} =
  \left\{
  \begin{array}{cc}
    \frac{1}{2 p_i\cdot Q_{i,s}}
  \frac{1}{1-\Sigma'(M_i^2)} + {\cal O}(\lambda)& \xi = 1 \text{ and } \Sigma_S(k^2) = \Sigma(k^2)\\
{\cal O}(\lambda) & \text{ otherwise }  \end{array} \right.
\end{equation}
where the physical and renormalized (complex) mass relate as $M_i^2 =
\tilde{M}_{R,i}^2+ \Sigma(M_i^2)$ for the boson in question. Thus
depending on how the unphysical scalar's self energy and the gauge
parameter relate to each other, the scalars will contribute at leading
power along with their related vector bosons, or not. We will
investigate this in more detail in the future. The simplest non-scalar case
to consider is that of a massive gauge boson. In an
$R_\xi$ gauge their numerator reads
\begin{equation}
  V^{\mu\nu}(q_i+K_{i,s},M_i) =- \eta^{\mu\nu} + (1-\xi) \frac{(q_i+K_{i,s})^\mu(q_i+K_{i,s})^\nu}{(q_i+K_{i,s})^2 - \xi M_i^2}\;.
\end{equation}
Using the momentum parametrization, eq.\,(3) in the main text, we have
\begin{equation}
  V^{\mu\nu}(q_i+K_{i,s},M_i) =- \eta^{\mu\nu} -\alpha^2 (1-\xi) \frac{(\Lambda p_i)^\mu (\Lambda p_i)^\nu}{2 p_i\cdot Q_{i,s} + (1 - \xi) M_i^2}
  = \sum_{\lambda}
  \epsilon_{\lambda,M}^\mu(\Lambda p_i,M_i)
  \epsilon_{\lambda,M}^{*,\nu}(\Lambda p_i,M_i) + {\cal O}(\lambda)
\end{equation}
where the physical polarization sum is $\sum_{\lambda}
\epsilon_{\lambda,M}^\mu(p_i,M_i)
\epsilon_{\lambda,M}^{*,\nu}(p_i,M_i) = -\eta^{\mu\nu} + p_i^\mu
p_i^\nu/M_i^2$. At tree level and $\xi=1$ this would not hold, but in
this case the Goldstones would contribute, as shown above. For $\xi\ne
1$ we obtain the physical polarization sum, and the Goldstones would
not contribute. The summed propagator is more complicated, however we
still find that the above relation holds with the expected additional
factor of the (complex) transverse self-energy of the respective
boson, as claimed at the beginning of this section. For Fermions the
propagator numerator is linear in the momentum and thus they trivially
satisfy our constraints at leading power. The last complicated case is
that of a massless gauge boson in physical gauge, for which we observe
that
\begin{equation}
  d^{\mu\nu}(q,n) = -\eta^{\mu\nu} + \frac{q^\mu n^\nu + n^\mu q^\nu}{n\cdot q}
\end{equation}
which clearly satisfies
\begin{equation}
  d^{\mu\nu}(q_i+K_{i,s},n) = d^{\mu\nu}(\Lambda p_i) + {\cal O}(\lambda) = \sum_\lambda \epsilon_\lambda^\mu(\Lambda p_i,n)
  \epsilon_{\lambda}^{*,\nu}(\Lambda p_i,n) + {\cal O}(\lambda)
\end{equation}
where $\epsilon(q,n)$ are the physical transverse polarization vectors
satisfying $\epsilon(q,n)\cdot q = \epsilon\cdot n = 0$ and
$\epsilon_\lambda(q,n)\cdot
\epsilon_\sigma(q,n)=-\delta_{\lambda\sigma}$. For a full propagator
we get the usual renormalization factor $1/(1-\Sigma_T(0))$ where
$\Sigma_T$ is the transverse part of the self energy, and the part
which is longitudinal to the gauge vector is again only contributing
at subleading power.
  
\end{document}